\documentclass[sigplan, screen]{acmart}

\AtBeginDocument{%
  \providecommand\BibTeX{{%
    \normalfont B\kern-0.5em{\scshape i\kern-0.25em b}\kern-0.8em\TeX}}}

\setcopyright{none}

\usepackage{wrapfig, amsthm, pict2e, booktabs, subcaption, subfiles, rotating, pifont, indentfirst, amsmath, mathbbol, amsfonts, mathrsfs, adjustbox, mathtools,booktabs, microtype, ebproof, enumitem, multirow, multicol, setspace, hyperref, xcolor, lineno, listings, stmaryrd, xspace, stackengine}
\usepackage{empheq}

\usepackage[tableposition=top]{caption}
\usepackage{tikz-cd}
\usetikzlibrary{decorations.pathmorphing}
\usepackage[ruled,linesnumbered,vlined]{algorithm2e}
\usepackage{graphicx}
\usepackage[frozencache,cachedir=.]{minted}
\usepackage{tikz}
\newcounter{markeq}
\usepackage{spacingtricks}

\usepackage[T1]{fontenc}

\usepackage{spacingtricks, commands, refinementtydef}

\begin{document}

\title{Type-Based Incorrectness Reasoning}

\author{Zhe Zhou}
\orcid{0000-0003-3900-7501}             %% \orcid is optional
 \affiliation{
   \institution{Purdue University}            %% \institution is required
  \country{}                    %% \country is recommended
}
\email{zhou956@purdue.edu}          %% \email is recommended

\author{Benjamin Delaware}
\orcid{0000-0002-1016-6261}
\affiliation{
  \institution{Purdue University}            %% \institution is required
  \country{}                    %% \country is recommended
}
\email{bendy@purdue.edu}

\author{Suresh Jagannathan}
\orcid{0000-0001-6871-2424}
\affiliation{
  \institution{Purdue University}            %% \institution is required
  \country{}                    %% \country is recommended
}
\email{suresh@cs.purdue.edu}


\renewcommand{\shortauthors}{Zhe Zhou, Benjamin Delaware, and Suresh Jagannathan}

\begin{abstract}
  A coverage type generalizes refinement types found in many
  functional languages with support for \emph{must}-style
  underapproximate reasoning.  Property-based testing frameworks are
  one particularly useful domain where such capabilities are useful as
  they allow us to verify the \emph{completeness}, as well as safety,
  of test generators.  There is a surprising connection between the
  kind of underapproximate reasoning coverage types offer and the
  style of reasoning enabled by recently proposed Incorrectness Logic
  frameworks.  In our presentation, we propose to explore this
  connection more deeply, identifying mechanisms that more
  systematically integrate incorrectness reasoning within an
  expressive refinement type system and the opportunities that such
  integration offers to functional programmers, program verifiers, and
  program analyzers and related tools.
\end{abstract}


%\keywords{TODO}

\maketitle

\section{Introduction}

In recent work, we proposed \emph{coverage refinement types}\cite{CT,
  CTArtifact} as a mechanism to automatically verify the
\emph{completeness} of test input generators used in property-based
testing (PBT)
frameworks~\cite{quickcheck-coverage-guided,claessen2011quickcheck}. An
input generator is complete with respect to some property if it is
capable of producing \emph{all} elements that satisfy that
property. Analogous to how Incorrectness Logic (IL)\cite{IL} precisely
captures the conditions that can trigger a program fault, our coverage
type system uses ``\emph{must}-style'' underapproximate reasoning
principles to characterize the possible outputs of a program. In our
system, properties are encoded as type qualifiers in the vein of
refinement type systems~\cite{JV21}. When a program type checks
against a coverage type, however, it is guaranteed to be able to
produce \emph{all} values satisfying the qualifier of that type, in
contrast to standard refinement type systems in which a refinement
constrains the set of values an expression \emph{may} produce.

% Underapproximate reasoning is conducted through both coverage type and incorrectness logic, each addressing distinct challenges. IL is dedicated to prove program indeed buggy, whereas coverage type focuses on the coverage completeness verification of test input generators that used in PBT. In general, the coverage type also aims to the bug finding, but in an indirectly way. Poirot and PBT treats the programs under test are totally black-box, and test it with test cases provided by a test generator that is well-typed by our coverage type system.

\begin{figure}[b!]
    \centering
    \includegraphics[width=240pt]{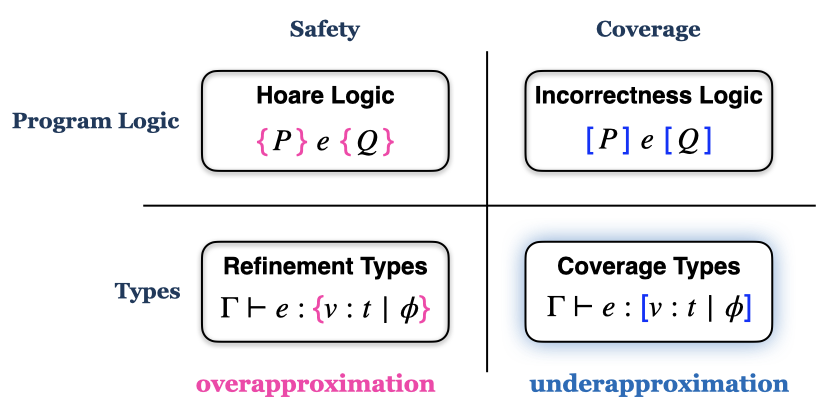}
    \caption{The full picture of over- and under-approximate reasoning in program logic and types.}
    \label{fig:comparison}
\end{figure}

Verifying completeness properties in the context of PBT is quite
different from the typical use case of IL, which has been primarily
studied in the context of bug finding~\cite{IL,ISL, ILRealBugs}. It is
interesting to ask if coverage types are sufficiently expressive to
also verify program incorrectness, similar to other IL
proposals. Our proposed talk will explore this question, showing how
coverage types can be used to reason about the incorrectness of OCaml
programs, using a small set of straightforward extensions to the
original source language and type system of \citet{CT}.

% Affirmatively, our coverage type system can perform analogous incorrectness verification on functional programs. This process utilizes IL-style assumptions and assertions, represented as straightforward syntactic sugar in the simply-typed lambda calculus (STLC). This method essentially offers a type-theoretic interpretation of incorrectness logic within functional languages.

% The remainder of the paper is structured as follows. Before we move on the type-based incorrectness verification, we provide a brief background of the coverage types and the corresponding comparison with the incorrectness logic (IL) in Section~\ref{sec:background}.
% Section~\ref{sec:overview} illustrate how the type-based incorrectness verification works. Finally, the discussion about the future works and conclusions are given in Section~\ref{sec:future}.

\section{Coverage Types and Incorrectness Logic}\label{sec:background}

Coverage types can be seen as a type-based interpretation of
Incorrectness Logic. As shown in \autoref{fig:comparison}, similar to
how an incorrectness triple $\utparenth{p}\,C\,\utparenth{q}$
reformulates a Hoare triple $\otparenth{p}\,C\,\otparenth{q}$ to
assert the existence of paths, coverage types ascribes a
\emph{must}-style interpretation over refinement type qualifiers,
rather than the \emph{may}-style interpretation used by a standard
refinement type system~\cite{JV21}.
% In the following, all programs discussed in this paper are written in simply typed lambada calculus (STLC) and OCaml code style, as in the original coverage type paper.
A standard refinement type $\rawnuot{b}{\phi}$ refines a basic type
$b$ with a logical formula, or \emph{qualifier} $\phi$, which states
that ``a program \emph{may} only reduce to a value $\vnu$ satisfying
$\phi$''. In our coverage type system, a base type $b$ can also be
qualified using a qualifier $\phi$, but $\phi$ now constrains the set
of values a program \emph{must} evaluate to. We denote these qualified
types using brackets $\rawnuut{b}{\phi}$ to emphasize that a coverage
type is interpreted differently than a standard refinement type an
application of the built-in random number generator
($\Code{int\_gen} : \Code{unit\sarr int}$), for example, has the
coverage type $\Code{int\_gen~()}: \nuut{int}{\top}$.

\begin{table}[t!]
\renewcommand{\arraystretch}{0.8}
\caption{\small Examples of overapproximate and underapproximate
  (coverage) typings. We use $\goodtype$ and $\badtype$ to indicate
  whether a term can or cannot be assigned the corresponding type,
  resp. }
\vspace*{-.1in}
\footnotesize
\begin{tabular}{l|lll}
\toprule
  $\Code{int\_gen\,()}$
  & $\goodtype \nuut{int}{\top}$
  & $\goodtype \nuut{int}{\nu = 1 \lor \nu = 2}$
  & $\goodtype \nuut{int}{\nu = 1}$ \\
  & $\goodtype \nuot{int}{\top}$
  & $\badtype \nuot{int}{\nu = 1 \lor \nu = 2}$
  & $\badtype \nuot{int}{\nu = 1}$ \\
\midrule
  $\Code{1}$
  & $\badtype \nuut{int}{\top}$
  & $\badtype \nuut{int}{\nu = 1 \lor \nu = 2}$
  & $\goodtype \nuut{int}{\nu = 1}$\\
  & $\goodtype \nuot{int}{\top}$
  & $\goodtype \nuot{int}{\nu = 1 \lor \nu = 2}$
  & $\goodtype \nuot{int}{\nu = 1}$ \\
\bottomrule
\end{tabular}
\label{tab:examples}
\vspace*{-.15in}
\end{table}

To illustrate the distinction between standard and coverage refinement
types, consider the combinations of expressions and types shown in
\autoref{tab:examples}. These examples illustrate that it is always
possible to strengthen the refinement qualifier used in an
underapproximate type and weaken such a qualifier in an
overapproximate type. A similar phenomena appears in IL's rule of
consequence, which inverts the direction of the implications on pre-
and postconditions in the overapproximate version of the rule.

\begin{table}[t!]
\renewcommand{\arraystretch}{0.8}
\caption{\small The semantics of type-based and program-logic-based
  assertions. We use the standard multi-step reduction relation
  ($\hookrightarrow^*$) and big-step relation relation ($\Downarrow$)
  to denote the operational semantics of the language under consideration.
% We use
% $s_{\Code{init}}$ and $p_{\Code{init}}$ to indicates the initial state (i.e., all variables has default value) and corresponding precondition. Then we have $\forall s. s \models p_{\Code{init}} \iff s = s_{\Code{init}}$, where $s \models p$ means a state $s$ makes the property $p$ hold.
}
\vspace*{-.1in}
\footnotesize
\begin{tabular}{l|l}
\toprule
  $\goodtype e : \rawnuot{b}{\phi}$
  & $\forall v{:}b. \mstep{e}{v} \impl \phi[\vnu \mapsto v]$ \\
\midrule
  $\goodtype e : \rawnuut{b}{\phi}$
  & $\forall v{:}b. \phi[\vnu \mapsto v] \impl \mstep{e}{v}$ \\
\midrule
  $\otparenth{\top}\,C\,\otparenth{q}$
  & $\forall s. (\exists s_0. (s_0, C)\Downarrow (s, \Code{SKIP})) \impl s \models q$ \\
\midrule
  $\utparenth{\top}\,C\,\utparenth{q}$
  & $\forall s. s \models q \impl (\exists s_0.(s_0, C) \Downarrow (s, \Code{SKIP}))$ \\
\bottomrule
\end{tabular}
\label{tab:comparison}
\vspace*{-.15in}
\end{table}

Table~\ref{tab:comparison} provides the semantics of type-based and
axiomatic assertions found in safety and incorrectness type
frameworks.  Note that both incorrectness logic and coverage types
invert the implication from the corresponding safety rule.  In other
words, incorrectness assertions imply the existence of a specific
execution path of the program, encoded as either a multi-step reduction chain
$\mstep{e}{v}$  or a big-step reduction relation $(s_0, C)\Downarrow (s, \Code{SKIP})$ from the
initial state $s_0$.

% Note that, a functional language like STLC don't have global variables (i.e., all variables are local), thus, the comparison only makes sense over close terms in both the in the functional and imperative languages.
% Following this setting, we assume the precondition in Hoare triple and incorrectness triple is $p_{\Code{init}}$, which indicates all variables in the abstract program state are initialized as default values. Then, it is clear that both coverage types and incorrectness logic inverts the implication between the specification (i.e., qualifier $\phi$ and postcondition $q$) and multi-step reduction relation (i.e.,$\mstep{}{}$).

\section{Type-Based Incorrectness Reasoning, By Example}
\label{sec:overview}

We now present, by way of examples, how we might augment Poirot, our
implementation of coverage types for OCaml, to support incorrectness
reasoning, more generally.

% Following our previous discussion, Poirot and PBT also aims bug catching but using an testing approach instead of verification like the IL. Thus, the type-based incorrectness reasoning simply adapts how testing framework using test input generators, but do static type check instead of dynamic execution on the programs under test.

\begin{example}[Incorrectness Reasoning]\label{ex:incor} Consider the
  following implementation of the example from Section 3.1
  of~\citet{IL}, rewritten in OCaml:
\begin{minted}[xleftmargin=10pt, numbersep=4pt, linenos = true, fontsize = \small, escapeinside=!!]{ocaml}
(* presumes : z==11 *)
if is_even x then
    if is_odd y then 42 else z
else z
(* achieves : 42 *)
\end{minted}

\noindent where the desired pre- and post-condition are shown in the
comments. In the setting of IL, the assignment of global variables
$\Code{x,y}$, and $\Code{z}$ are provided by the ambient program
state. We can achieve a similar result here by using test input
generators consistent with the precondition (i.e., $\Code{int\_gen}$
and $\Code{int\_range}$) to initialize these variables.

\begin{minted}[xleftmargin=10pt, numbersep=4pt, linenos = true, fontsize = \small, escapeinside=!!]{ocaml}
let (x: int) = int_gen () in
let (y: int) = int_gen () in
let (z: int) = int_range 11 11 in
if is_even x then if is_odd y then 42 else z else z
\end{minted}

% where the assignment of variable $\Code{x}$ and $\Code{y}$ are provided the generator $\Code{int\_gen}$ since the precondition has no constant over them; on the other hand, the $\Code{int\_range}$ is test generator returns random integers in the given inclusive range.

\noindent We can apply coverage typechecking to ascertain if there
exists a path to the expected result value $42$ by checking if this
program has the coverage type $\nuut{int}{\vnu = 42}$.  After
translating the precondition into applications of input generators,
Poirot is able to prove the program above can indeed achieve result
value $42$. Since we are working in a pure language, our postcondition
does not need to refer to the variable $\Code{x}$ and $\Code{y}$, which
avoids the need to have the user user supply a more precise
postcondition (i.e., $\Code{x\ is\ even} \land \Code{y\ is\ odd}$) as
in IL.
\end{example}

\paragraph{Intuition} This example illustrates how can perform
incorrectness reasoning with coverage types.  After modifying the
original program by adding a random initialization procedure
consistent with an underapproximate precondition at the beginning of
the code fragment, we can then ask if it can cover all values
consistent with the underapproximate postcondtion. For example, the
program under test in this example takes the results of generators
(e.g., {\small$\Code{int\_range\ 11\ 11}$}), and can be treated as a
generator that covers the value $42$.

Following this intuition, we new describe a more systematic approach
to typed-based incorrectness reasoning in an expressive functional
language like OCaml, focusing in particular on the assumptions and
assertions from imperative language that don't exist in a pure
functional language.

\paragraph{Assumptions} We have informally seen how a coverage type
system can verify the incorrectness of a given program, provided the
programmer has inserted appropriate in generators (e.g.,
$\Code{int\_range}\ 11\ 11$) to mark the space of inputs consistent
with a desired precondition. To improve usability and readability,
however, we can instead introduce dedicated syntax to express our
assumption on the existence of such a generator.  We do so via the
term $\Code{assume}\ \rawnuut{b}{\phi} \equiv \Code{f\ ()}$, which
assumes the existence of some function $\Code{f}$ that has been
type-checked against the function type
$\Code{unit}\sarr\rawnuut{b}{\phi}$; this term represents some unknown
input generator that covers all values in qualifier $\phi$.

\begin{example}[Assumptions]\label{ex:assume} Example~\ref{ex:incor} can be rewritten
  using ``$\Code{assume}$'' as follows:
\begin{minted}[xleftmargin=10pt, numbersep=4pt, linenos = true, fontsize = \small, escapeinside=!!]{ocaml}
let (x: int) = assume !$\nuut{int}{\top}$! in
let (y: int) = assume !$\nuut{int}{\top}$! in
let (z: int) = assume !$\nuut{int}{\vnu = 11}$! in
if is_even x then if is_odd y then 42 else z else z
\end{minted}
\end{example}

\paragraph{Errors} One interesting feature of IL is that its
postconditions are able to track both correct
($\textcolor{DeepGreen}{[ok: r]}$) and erroneous paths
($\textcolor{red}{[err: r]}$).  As we will now demonstrate, this
feature can be captured by an error monad that is similar to the
``result'' type in OCaml
\begin{minted}[xleftmargin=10pt, numbersep=4pt, linenos = true, fontsize = \small, escapeinside=!!]{ocaml}
type 'a monad = bool * 'a
(* monad bind *)
let (let*) (x: bool * 'a) (f: 'a -> bool * 'a) =
    match x with
    | (true, x') -> (true, f x') (* !$ok$! in IL *)
    | (false, x') -> (false, x') (* !$err$! in IL *)
\end{minted}

\noindent where the monad carries an additional boolean value.  Here,
$\Code{true}$ indicates if the execution terminates normally (i.e.,
$\textcolor{DeepGreen}{ok}$) and $\Code{false}$ when it fails (i.e.,
$\textcolor{red}{err}$). As a result, the bind operation only applies
the input function when the execution terminates normally (line $5$).

\begin{example}[Errors]\label{ex:error} Consider the following
  program, adapted from Fig.~8 in~\citet{IL}:
\begin{minted}[xleftmargin=10pt, numbersep=4pt, linenos = true, fontsize = \small, escapeinside=!!]{ocaml}
let foo (x: int) : int monad =
    if is_even x then (false, x)
    else if bool_gen () then (true, x)
    else (false, x)
(*  achieves [err: x is odd],[ ok: x is odd] *)
\end{minted}

\noindent where only the second branch terminates correctly (line $3$),
returning a pair whose first element is
$\Code{true}$. Then, the IL-style postcondition shown on line $5$ can
be encoded as the following coverage type {\small\begin{align*}
  \nuut{bool \times int}{\neg \I{fst}(\vnu) \land
    \I{odd}(\I{snd}(\vnu)) \lor \I{fst}(\vnu) \land
    \I{odd}(\I{snd}(\vnu))}
\end{align*}}\noindent
where the operators $\I{fst}$ and $\I{snd}$ indicate the first and
second element of a pair, and $\I{odd}$ and $\I{even}$ returns true iff
input integer is odd or even number.
\end{example}

\paragraph{Assertions} With the error monad, assertions can
straightforwardly be encoded as a function application that returns a
unit value when a given property $\phi$ holds.
{\small
  \begin{align*}
    &\Code{assert\ }\phi \equiv \Code{g}\ () \\&\quad \text{ where }
    \Code{g} : \Code{unit}\sarr\nuut{bool \times unit}{\neg
    \I{fst}(\vnu) \land \neg \phi \lor \I{fst}(\vnu) \land \phi }
  \end{align*}}\noindent

\begin{example}[Assertions]\label{ex:assert} The client program from
  Fig.8 in~\citet{IL} can be rewritten as:
\begin{minted}[xleftmargin=10pt, numbersep=4pt, linenos = true, fontsize = \small, escapeinside=!!]{ocaml}
let flaky_client () : int monad =
    let x = 3 in
    let* () = foo 3 in
    let y = x + 2 in
    let* () = assert (!$x = 4$!) in
    (true, x)
(* achieves : [err: x==3 || x==5] *)
\end{minted}

\noindent where we use the bind operation ($\Code{let^*}$) instead of
a normal let-expression on line $9$ and $11$, to account for the
possibility that the both the use of $\Code{foo}$ and the
$\Code{assert}$ may fail. The $\Code{flaky\_client}$ function can then
be type-checked against the coverage type
{ \small
  \begin{align*}
    \nuut{bool \times int}{\neg \I{fst}(\vnu) \land (\I{snd}(\vnu) = 3
      \lor \I{snd}(\vnu) = 5)}
  \end{align*}}\noindent
whose qualifier matches the IL-style postcondition on line $7$.
\end{example}

% \subfile{sections/language}
\section{Discussion and Future Work}\label{sec:future}

These examples establish a number of tantalizing links between
coverage types and IL. There are several interesting future directions
we hope to explore to flesh out these connections further; we list
just one of them below, and plan to discuss these links in more detail
at our presentation, should this proposal be accepted.

\paragraph{Underapproximate parameter types} Reconsider
Example~\ref{ex:assume}, which used an \Code{assume} statement to
ascribe the variables $\Code{x}$, $\Code{y}$, and $\Code{z}$ coverage
types consistent with the values expected by the precondition. A more
general way to think about computations like these is to abstract them
as functions with underapproximate parameter types (e.g.,
$(\Code{x}:\nuut{int}{\top})$): \begin{minted}[xleftmargin=10pt,
  numbersep=4pt, linenos = true, fontsize = \small,
  escapeinside=!!]{ocaml}
  let f (x:!$\nuut{int}{\top}$!) (y:!$\nuut{int}{\top}$!)
  (z:!$\nuut{int}{\vnu = 11}$!) =
    if is_even x then
        if is_odd y then 42 else z
    else z
\end{minted}

Poirot currently does not support such a feature, however. In our
current system, function types are restricted to have
\emph{overapproximate} parameter types (i.e.,
$x{:}\rawnuot{b}{\phi}\sarr\tau$) instead of an underapproximate ones
(i.e., $x{:}\rawnuut{b}{\phi}\sarr\tau$). This design choice aligns
with how test input generators are typically used, which require a
safety check instead of a coverage completeness check on their
arguments. As an example, a sensible sized list generator
($\Code{sized\_list\_gen}: \Code{int \sarr int\ list}$), will allow
users to choose any integer that is greater than or equal to $0$ as a
valid (and \emph{safe}) argument.

Extending Poirot to support underapproximate parameter types
introduces interesting challenges. As an example, consider the
function type
$\tau \equiv \nuut{int}{\top}\sarr\nuut{int}{1 \leq \vnu \leq 2}$ with
underapproximate parameter types, and the following pair of valid
implementations of this type:
\begin{minted}[xleftmargin=10pt, numbersep=4pt, linenos = true, fontsize = \small, escapeinside=!!]{ocaml}
let imp1 (x: int) = if x > 0 then 1 else 2
let imp2 (x: int) = if x > 0 then 2 else 1
\end{minted}
\noindent The following program that uses two functions
$\Code{f}$ and $\Code{g}$ of type $\tau$:
\begin{minted}[xleftmargin=10pt, numbersep=4pt, linenos = true, fontsize = \small, escapeinside=!!]{ocaml}
  let client =
  let (x: int) = int_gen () in (f x) - (g x) in x
\end{minted}

\noindent There are now two scenarios, depending on how $\Code{f}$ and
$\Code{g}$ are instantiated:
\begin{enumerate}
\item The body of both $\Code{f}$ and $\Code{g}$ is $\Code{imp1}$. In
  this scenario, $\Code{f\ x}$ would \emph{always} equal
  $\Code{g\ x}$, so $\Code{client}$ can only cover (aka reach) the
  value $0$ (i.e., $\nuut{int}{\vnu = 0}$).
    \item The bodies of $\Code{f}$ and $\Code{g}$ are $\Code{imp1}$ and  $\Code{imp2}$, respectively. Then, $\Code{f\ x}$ is \emph{never} equal to $\Code{g\ x}$, so $\Code{client}$ can cover (aka reach) both $1$ and ${-}1$; i.e., it has the coverage type $\nuut{int}{\vnu = 1 \lor \vnu = {-}1}$.
\end{enumerate}
Note that we cannot distinguish between these two cases based solely
on the signatures of $\Code{f}$ and $\Code{g}$. More problematically,
the types that can be assigned to $\Code{client}$ in these scenarios
have disjoint qualifiers, so their intersection is empty. 
% ($\nuut{int}{(\vnu = 1 \lor \vnu = {-}1) \land \vnu = 0}$)
The issue speaks to the deficiencies of using underapproximate
parameters when verifying client-side applications. This represents a
precision trade-off between a library and its client, that
necessitates further exploration.

A similar issue is discussed in Section 6.4 of \citet{IL}, which
proposes introducing a ``logical variable'' $m$ to assign a more
precise type to $\Code{imp_1}$ {\small
  \begin{align*}
    \nuut{bool}{\vnu = m}\sarr\nuut{int}{ m > 0 \land \vnu = 1 \lor
      m\leq 0 \land \vnu = 2}
  \end{align*}
}\noindent Such logical variables would need to be automatically
instantiated during type checking, an intriguing and challenging
direction we would like to explore in the future.

\bibliographystyle{ACM-Reference-Format}
\bibliography{bibliography}

\appendix

\end{document}